\documentclass[11pt]{article}\usepackage[hyperfootnotes=false]{hyperref}
   \usepackage{epsfig}
      \usepackage{amsmath}
      \usepackage{amssymb}
  \usepackage{graphicx}
  \newlength{\abstractwidth}
  \usepackage{graphicx}                            
  \usepackage{amsmath} 
  \usepackage{sidecap} 
  
  \usepackage{hyperref}
  \usepackage{amsthm}

  \renewcommand{\thefootnote}{\fnsymbol{footnote}}
  \renewcommand{\thanks}[1]{\footnote{#1}} 
  \newcommand{\starttext}{
  \setcounter{footnote}{0}
  \renewcommand{\thefootnote}{\arabic{footnote}}}
  \renewcommand{\theequation}{\thesection.\arabic{equation}}
  \newcommand{\be}{\begin{equation}}
  \newcommand{\bea}{\begin{align}}
  \newcommand{\eea}{\end{align}}
  \newcommand{\beq}{\begin{equation}}
  \newcommand{\ee}{\end{equation}}
  \newcommand{\eeq}{\end{equation}}

  \newcommand{\ket}[1]{\left|#1\right\rangle}

  \def\ba{\begin{eqnarray}}
  \def\ea{\end{eqnarray}}

  \def\12{{1 \over 2}}

 \def\simleq{\; \raise0.3ex\hbox{$<$\kern-0.75em
      \raise-1.1ex\hbox{$\sim$}}\; }
 \def\simgeq{\; \raise0.3ex\hbox{$>$\kern-0.75em
      \raise-1.1ex\hbox{$\sim$}}\; }

\def\O2{\Omega_2}

\def\bi{\begin{itemize}}
  \def\ei{\end{itemize}}
\def\sc{\setcounter{equation}{0}}

\def\tg{\tilde{\gamma}}
\def\ser{S_{ER}}
\def\ier{I_{ER}}

\def\ereqepr{$\text{ER}=\text{EPR}$}

\begin{document}

  \renewcommand{\theequation}{\thesection.\arabic{equation}}

\begin{titlepage}
  \rightline{}
  \bigskip

  \bigskip\bigskip\bigskip\bigskip

    \bigskip
\centerline{\Large \bf {Are entangled particles connected by wormholes?}}
\vspace{0.3cm}
\centerline{\Large \bf {Support for the \ereqepr\ conjecture from entropy inequalities
}}
    \bigskip

\begin{center}

\bf {{Hrant Gharibyan$^1$ and  Robert F. Penna$^{1,2}$}}
\bigskip \rm
\bigskip

$^1$Department of Physics, and Kavli Institute for Astrophysics and Space Research,
Massachusetts Institute of Technology, Cambridge, MA 02139, USA\\
$^2$Harvard-Smithsonian Center for Astrophysics, 
Cambridge, MA 02138, USA
\\ ~~
\\

\rm

\bigskip
\bigskip

\vspace{1cm}
  \end{center}

  \bigskip\bigskip

 \bigskip\bigskip
  \begin{abstract}

If spacetime is built out of quantum bits, does the shape of space depend on how the bits are entangled?  The \ereqepr\ conjecture relates the entanglement entropy of a collection of black holes to the cross sectional area of Einstein-Rosen (ER) bridges (or wormholes) connecting them.  We show that the geometrical entropy of classical ER bridges satisfies the subadditivity, triangle, strong subadditivity, and CLW inequalities.    These are nontrivial properties of entanglement entropy, so this is evidence for \ereqepr.    We further show that the entanglement entropy associated to classical ER bridges has nonpositive interaction information.   This is not a property of entanglement entropy, in general.  For example, the entangled four qubit pure state $\ket{GHZ_4}=(\ket{0000}+\ket{1111})/{\sqrt{2}}$ has positive interaction information, so this state cannot be described by a classical ER bridge.  Large black holes with massive amounts of entanglement between them can fail to have a classical ER bridge if they are built out of $\ket{GHZ_4}$ states.  States with nonpositive interaction information are called monogamous.  We conclude that classical ER bridges require monogamous EPR correlations.  

 \medskip
  \noindent
  \end{abstract}

  \end{titlepage}

    \starttext 
    \setcounter{footnote}{0}

  \sc
  
\section{Introduction}

It has been suggested that spacetime has an underlying, quantum information theoretic description \cite{2010GReGr..42.2323V}.  The \ereqepr\ conjecture \cite{2013arXiv1306.0533M}  is a specific realization of this proposal.  The conjecture says that the quantum degrees of freedom corresponding to black holes connected by an Einstein-Rosen (ER) bridge \cite{1935PhRv...48...73E}  are entangled.   The entanglement entropy is related to the cross sectional area of the bridge.  The conjecture also claims the converse, viz. that all entangled states are connected by an ER bridge of some sort, although in many cases the bridge is a highly quantum object with no independent definition.  The conjecture is summarized as \ereqepr, where EPR stands for Einstein, Podolsky, and Rosen \cite{1935PhRv...47..777E} and represents entanglement.

There is indirect support for the conjecture \cite{2013arXiv1306.0533M}.   For instance, it is a curious coincidence that despite the nonlocal nature of ER bridges and EPR correlations, they both conspire to prevent superluminal signals.  On the ER side, this is a consequence of the topological censorship theorems \cite{1962PhRv..128..919F,1993PhRvL..71.1486F,1999PhRvD..60j4039G}, which forbid traversable wormholes for reasonable energy conditions.  If traversable wormholes were possible, the \ereqepr\ conjecture would be wrong.  Other evidence comes from finding examples.   A certain entangled pair has been realized as an ER bridge using AdS/CFT  \cite{2013arXiv1307.1132J,2013arXiv1307.6850S}.   

The conjecture has invited criticism.  \ereqepr\ is partly based on intuition from the dual CFT description of eternal black holes in AdS, but this is a special state and its behavior might not be representative of generic entangled states \cite{2013arXiv1307.4706M}.  It is not clear how differences between distinct entangled states with the same entanglement entropy are to be realized in the ER bridge description \cite{2013arXiv1307.1604N}.  It will probably be some time before the conjecture is resolved one way or the other.

In this paper we consider a different kind of test of the conjecture. We show that the geometrical notion of entropy associated to ER bridges obeys the same inequalities as ordinary entanglement entropy: subadditivity, the triangle inequality, strong subadditivity, and a set of inequalities discovered by Cadney, Linden and Winter (CLW) \cite{2005CMaPh.259..129L,2011arXiv1107.0624C}.   This is evidence for the \ereqepr\ conjecture.

We also describe a restriction the conjecture imposes on the entangled degrees of freedom underlying classical ER bridges.  The entanglement must have nonpositive interaction information.  This is not a general property of quantum systems.  For instance, the four qubit pure state $\ket{GHZ_4}=(\ket{0000}+\ket{1111})/{\sqrt{2}}$ has positive interaction information.  Macroscopic black holes can be built out of many copies of $\ket{GHZ_4}$ which do not have classical ER bridges.  This serves to emphasize that the ER bridge depends on the pattern of entanglement and not just the total amount of entanglement.  

The \ereqepr\ conjecture is closely related to the holographic entanglement entropy (HEE) proposal for asymptotically AdS spaces.  This proposal relates the entanglement between regions on the AdS boundary to the area of a minimal surface in the bulk \cite{2006PhRvL..96r1602R,2006JHEP...08..045R}.  
Part of the evidence for the HEE proposal comes from the fact that it reproduces subadditivity, strong subadditivity \cite{2007PhRvD..76j6013H}, and the CLW inequalities \cite{2013PhRvD..87d6003H}.   
Given the similarities between HEE and the \ereqepr\ conjecture, it is not too surprising that they give the same entropy inequalities.  The proofs are related, but the setup is different.  Holographic entanglement entropy is defined for boundary regions of AdS spaces, whereas the \ereqepr\ conjecture applies to black holes in all spacetimes.  This includes astrophysical black holes, in principle.  So the \ereqepr\ conjecture might be of interest to astrophysicists and cosmologists who are unfamiliar with AdS.  This forms part of the motivation for the present work.

Our paper is organized as follows.   We review the \ereqepr\ conjecture in section \ref{sec:review}, prove the entropy inequalities in section \ref{sec:inequalities},  discuss implications for the conjecture in section \ref{sec:discuss}, and summarize our results in section \ref{sec:conclude} .

\section{The \ereqepr\ conjecture}
\label{sec:review}

\begin{figure}[!ht]
\begin{center}
\includegraphics[scale=1]{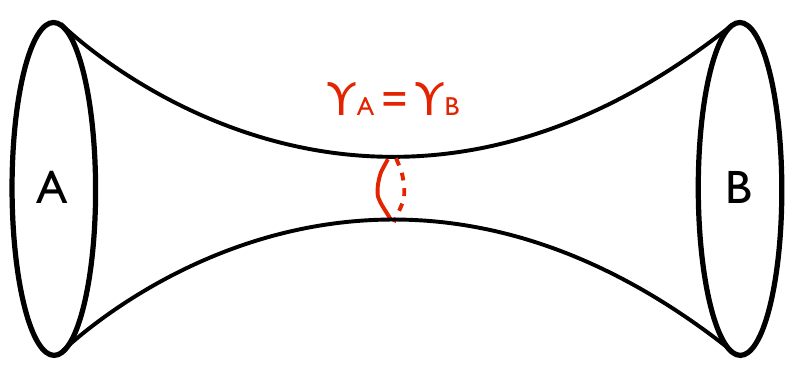}
\caption{A constant time slice of 2+1 dimensional spacetime is a two dimensional surface.  We assume the spacetime is static so the slice is well defined.  The black hole horizons are circles $A$ and $B$.  They are joined by an ER bridge.  The smallest cut separating the ER bridge into a piece containing $A$ and a piece containing $B$ is a closed curve, $\gamma_A$.  The length of $\gamma_A$ is $a(\gamma_A)=\ser(A)$.  For the simple geometry pictured here, in which $A$ and $B$ are not joined to any other black holes by ER bridges, $\ser(A)=\ser(B)$.}
\label{fig:AB}
\end{center}
\end{figure}

The \ereqepr\ conjecture relates the entanglement entropy of black holes connected by an ER bridge to the cross sectional area of the bridge.  We assume spacetime is static and work on a constant time slice.  Suppose black hole $A$ is joined to some other black holes by an ER bridge.  We cut the ER bridge into two pieces, such that one piece contains $A$ and the other piece contains the remaining black holes.  Let $\gamma_A$ be the cut with the smallest cross sectional area and let $a(\gamma_A)$ be its area.  Then the entanglement entropy of $A$ is conjectured to be
\beq\label{eq:def1}
\ser(A) \equiv a(\gamma_A).
\eeq
It follows that if two black holes $A$ and $B$ are not connected to any other black holes by ER bridges, then $\ser(A)=\ser(B)$.  Figure \ref{fig:AB} gives a 2+1 dimensional example.

Similarly, if $\gamma_{AB}$ is the area minimizing cut that separates black holes $A$ and $B$ from all other black holes, then 
\beq\label{eq:def2}
\ser(AB) \equiv a(\gamma_{AB}).
\eeq
This is zero if $A$ and $B$ are not connected to any other black holes by an ER bridge.  In this case the quantum degrees of freedom underlying the joint system $AB$ are conjectured to be in a pure state.

Definitions \eqref{eq:def1} and \eqref{eq:def2} are closely related to the notion of HEE in asymptotically AdS spaces \cite{2006PhRvL..96r1602R,2006JHEP...08..045R}.  The latter entropy is also defined as the area of a minimal surface.  There are two notable differences.  First, the minimal surface that defines the HEE of a region $A$ is required to be anchored on the boundary of $A$, while the cuts through the ER bridge do not meet $A$ in general (see \cite{2007JHEP...07..062H} for a further discussion of this difference).  Second, HEE is generally infinite and must be regularized, while definitions \eqref{eq:def1} and \eqref{eq:def2} are finite.  

It is not at all apparent from definitions \eqref{eq:def1} and \eqref{eq:def2} that the entropy so defined obeys the usual inequalities: subadditivity, the triangle inequality, strong subadditivity, and the CLW inequalities.  If the inequalities failed, then the \ereqepr\ conjecture would be wrong.  In the next section, we show that the inequalities are satisfied for static, classical ER bridges.  This supports the \ereqepr\ conjecture in this case.  Definitions \eqref{eq:def1} and \eqref{eq:def2} could be extended to non-static spacetimes using an analogue of the covariant holographic entanglement entropy (CHEE) proposal for AdS \cite{2007JHEP...07..062H}.  We will not consider that extension here, but see \cite{2012JHEP...06..081C,2012arXiv1211.3494W} for a discussion of the entropy inequalities satisfied by CHEE.  It is not possible to check the entropy inequalities for quantum ER bridges as these have yet to be defined (independently of the \ereqepr\ conjecture).

 \section{Entropy inequalities}
\label{sec:inequalities}

\subsection{Subadditivity}
\label{sec:subadd}
Consider two black holes, $A$ and $B$, connected by a static, classical ER bridge (in any spacetime dimension).  They may also be connected to other black holes by classical ER bridges.  We claim
\beq\label{eq:str}
\ser(AB) \leq \ser(A) + \ser(B).
\eeq
This inequality is called subadditivity. Subadditivity is a property of entanglement entropy, so the \ereqepr\ conjecture would be wrong if equation \eqref{eq:str} failed.  

By definition $\ser(A)=a(\gamma_A)$, where $\gamma_A$ is the minimal cut through the ER bridge that divides it into two pieces such that one piece contains $A$ and the other piece contains all other black holes.  The cuts $\gamma_B$ and $\gamma_{AB}$ are defined similarly.
\begin{figure}[!ht]
\begin{center}
\includegraphics[scale=1]{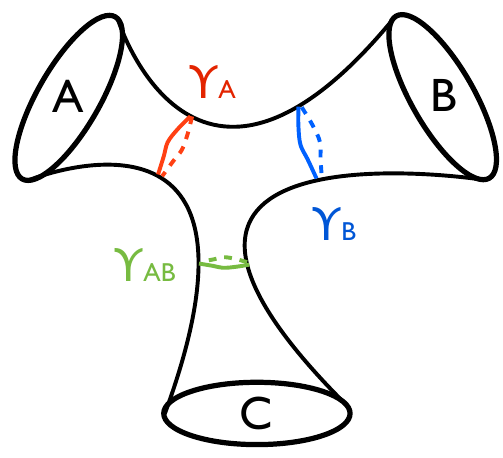}
\caption{An ER bridge on a constant-time slice of 2+1 dimensional spacetime is a two-dimensional surface.  (We assume spacetime is static, so the slice is well defined).  The black hole horizons are circles $A$, $B$, and $C$, and the minimum area cuts through the ER bridge are closed curves $\gamma_A$, $\gamma_B$, and $\gamma_{AB}$. The ER bridge pictured here has a simple geometry, but the proofs apply to all classical ER bridges.}
\label{fig:ABC}
\end{center}
\end{figure}

Now the combined cut, $\gamma_{A} \cup \gamma_{B}$, separates the ER bridge into a piece containing $AB$ and a piece containing any other black holes joined to AB by a classical ER bridge.  See Figure \ref{fig:ABC} for an example in 2+1 dimensions.  It does not generally have the minimum area of such a cut, but it gives an upper bound for $\ser(AB)$: 
\beq\label{eq:str1}
a(\gamma_A\cup \gamma_B)\geq a(\gamma_{AB}) = \ser(AB).
\eeq

On the other hand, the area of the combined cut is less than or equal to the combined areas of the separate cuts,
\beq\label{eq:str2}
a(\gamma_{A} \cup \gamma_{B}) \leq a(\gamma_{A}) + a(\gamma_{B})=\ser(A)+\ser(B),
\eeq 
where equality is obtained when $\gamma_A$ and $\gamma_B$ have no intersection.  Equations \eqref{eq:str1} and \eqref{eq:str2} together give subadditivity, \eqref{eq:str}.

\subsection{Triangle inequality}

Our next claim is that the ER bridge between $A$ and $B$ satisfies the triangle inequality:
\beq\label{eq:triangle}
|\ser(A)-\ser(B)| \leq \ser(AB).
\eeq
The triangle inequality is a standard property of entanglement entropy, so equation \eqref{eq:triangle} is another important check of the \ereqepr\ conjecture.

The combined cut $\gamma_B \cup \gamma_{AB}$ separates the ER bridge into a piece containing $A$ and a piece containing the remaining black holes. (See the 2+1 dimensional example in Figure \ref{fig:ABC}.)  It does not necessarily have the minimum area of such a cut, but it gives an upper bound for $\ser(A)$:
\beq\label{eq:tri1}
a(\gamma_B \cup \gamma_{AB}) \geq a(\gamma_A) = \ser(A).
\eeq

The area of the combined cut is less than or equal to the combined areas of the cuts, so
\beq\label{eq:tri2}
a(\gamma_B \cup \gamma_{AB}) \leq a(\gamma_B) + a(\gamma_{AB}) = \ser(B) + \ser(AB),
\eeq
where equality is obtained if $\gamma_B$ and $\gamma_{AB}$ have no intersection.
Equations \eqref{eq:tri1} and \eqref{eq:tri2} together imply
\beq
\ser(A) \leq \ser(B) + \ser(AB).
\eeq
We may assume $\ser(A) > \ser(B)$ without loss of generality and this immediately gives the triangle inequality \eqref{eq:triangle}, as desired.

\subsection{Strong subadditivity}

The next inequality we consider is strong subadditivity:
\beq\label{eq:strong}
\ser(A) + \ser(C) \leq \ser(AB) + \ser(BC).
\eeq
This offers the best support for \ereqepr\ of the three inequalities we have considered so far, as it is a significantly deeper property of entanglement entropy than either subadditivity or the triangle inequality.  Its depth can be seen from the difficulty of proving strong subadditivity on the EPR side for entanglement entropy \cite{1970CMaPh..18..160A,1973JMP....14.1938L}.

Strong subadditivity involves three black holes, $A$, $B$, and $C$, and the proof is a more elaborate version of our earlier arguments for subadditivity and the triangle inequality.  We consider the area minimizing cuts $\gamma_{AB}$ and $\gamma_{BC}$ whose areas are $\ser(AB)$ and $\ser(BC)$.  From these we define new cuts, $\tilde{\gamma}_{A}$ and $\tilde{\gamma}_C$, which split the ER bridge into pieces containing only $A$ and only $C$, respectively.  We show that the total area $a(\tilde{\gamma}_{A})+a(\tilde{\gamma}_C)$ is  bounded below by $\ser(A)+\ser(C)$ and above by $\ser(AB) + \ser(BC)$.  This will suffice to prove strong subadditivity.  The key step is defining $\tilde{\gamma}_{A}$ and $\tilde{\gamma}_C$.

The area minimizing cut $\gamma_{AB}$ splits the ER bridge into two pieces, one containing black holes $A$ and $B$ and the other containing all remaining black holes.  Let these two pieces be $ab$ and $cd$.   Similarly,  let $\gamma_{BC}$ split the ER bridge into two pieces $bc$ and $ad$.  

\begin{figure}[!ht]
\begin{center}
\includegraphics[scale=1]{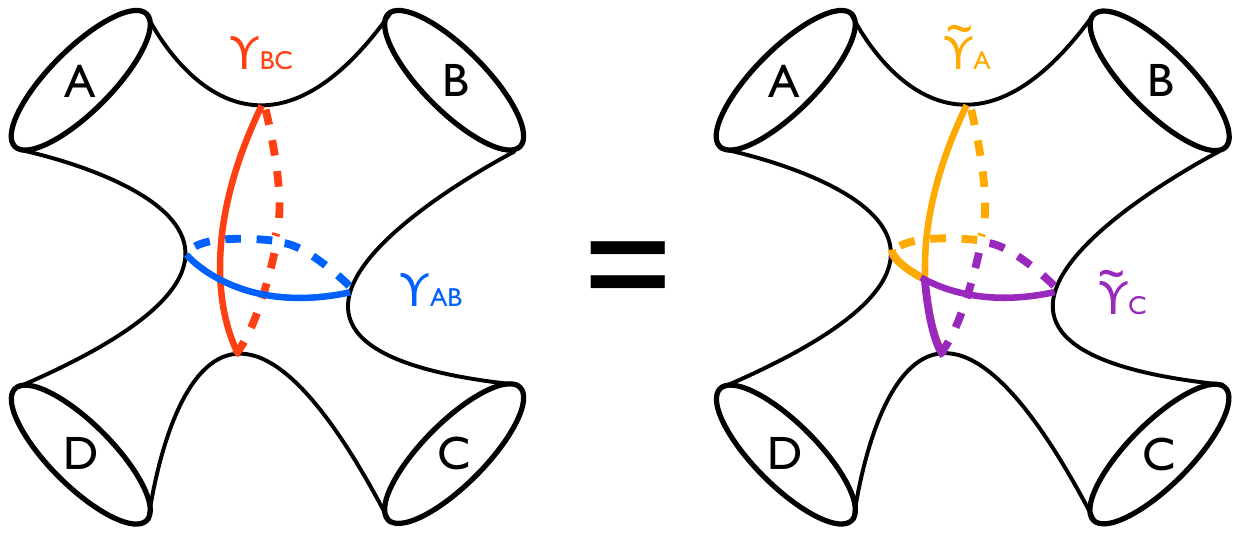}
\caption{As in Figure \ref{fig:ABC}, except now with four black holes, $A$, $B$, $C$, and $D$.  Starting from the cuts $\gamma_{AB}$ and $\gamma_{BC}$, we construct new cuts $\tilde{\gamma}_A$ and $\tilde{\gamma}_C$ (equations \ref{eq:strongcut1}-\ref{eq:strongcut2}).  Strong subaddivity is proved by bounding the sizes of $\tilde{\gamma}_A$ and $\tilde{\gamma}_C$.}
\label{fig:ABCD}
\end{center}
\end{figure}

The combinations we need are
\begin{align}
\tilde{\gamma}_A &\equiv \gamma_{AB}\vert_{ad}\cup\gamma_{BC}\vert_{ab},\label{eq:strongcut1}\\
\tilde{\gamma}_C &\equiv \gamma_{AB}\vert_{bc}\cup\gamma_{BC}\vert_{cd},\label{eq:strongcut2}
\end{align}
where $\gamma_{AB}\vert_{ad}$ is the restriction of $\gamma_{AB}$ to $ad$.  The new cut $\tilde{\gamma}_A$ splits the ER bridge into a piece containing black hole $A$ and a piece containing all remaining black holes, while $\tilde{\gamma}_C$ does the same with respect to black hole $C$.   Figure \ref{fig:ABCD} gives an example in 2+1 dimensions, involving four black holes connected by  a two-dimensional ER bridge. 

For all classical ER bridges in all dimensions, the total area $a(\tilde{\gamma}_A)+a(\tilde{\gamma}_C)$ is bounded below by the total area of the minimizing cuts:
\beq\label{eq:sublower}
a(\tilde{\gamma}_A) + a(\tilde{\gamma}_C) \geq a(\gamma_A) + a(\gamma_C) = \ser(A) +\ser(C).
\eeq
We also have the upper bound
\begin{align}
a(\tilde{\gamma}_A) + a(\tilde{\gamma}_C) \notag
&\leq a(\gamma_{AB}\vert_{ad})+a(\gamma_{BC}\vert_{ab}) + a(\gamma_{AB}\vert_{bc})+a(\gamma_{BC}\vert_{cd})\label{eq:subupper}\\
&= \ser(AB) + \ser(BC). 
\end{align}
Combining equations \eqref{eq:sublower} and \eqref{eq:subupper} gives strong subadditivity, equation \eqref{eq:strong}, as desired.

\subsection{Monogamy}

Finally, we will show that three black holes $A$, $B$, and $C$, connected by a static, classical ER bridge satisfy
\begin{align}\label{eq:mono}
\ser(A) + \ser(B) + \ser(C) + \ser(ABC) \notag\\
\leq  \ser(AB) + \ser(BC) + \ser(AC).
\end{align}
Systems obeying this inequality are called monogamous.  Not all entangled systems are monogamous in this sense, so equation \eqref{eq:mono} restricts the set of states that can describe static, classical ER bridges.  We return to the implications of this fact for \ereqepr\ in section \ref{sec:discuss}.

\begin{figure}[!ht]
\begin{center}
\includegraphics[scale=1]{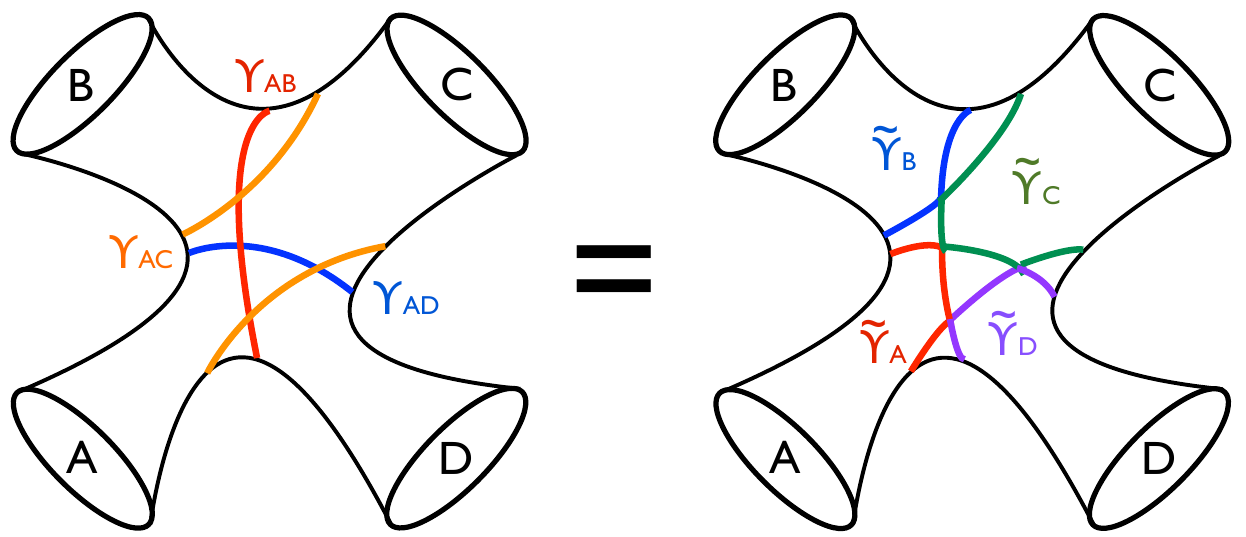}
\caption{A constant time slice of 2+1 dimensional spacetime is a two dimensional surface.  Black holes $A$, $B$, $C$, and $D$ (black circles) are connected by a two dimensional ER bridge.  Cuts through the bridge are rearranged to give new cuts, $\tg_A$, $\tg_B$, $\tg_C$, and $\tg_D$, each of which cuts out a single black hole (see equations \ref{eq:monocut1}-\ref{eq:monocut4}).  Monogamy is proved by bounding the sizes of the cuts.}
\label{fig:monogamy}
\end{center}
\end{figure}

Let $D$ be all other black holes connected to $A$, $B$, and $C$ by classical ER bridges ($D$ is not necessarily a single black hole).  We define
\ba 
\tg_A &=&\gamma_{AB}\vert_{ac \cap ad} \cup \gamma_{AC}\vert_{ad \cap ab} \cup \gamma_{AD}\vert_{ab \cap ac},
\label{eq:monocut1}  \\
\tg_B &=&\gamma_{BA}\vert_{bc \cap bd} \cup \gamma_{BC}\vert_{bd \cap ba} \cup \gamma_{BD}\vert_{ba \cap bc},\\
\tg_C &=&\gamma_{CA}\vert_{cb \cap cd} \cup \gamma_{CB}\vert_{cd \cap ca} \cup \gamma_{CD}\vert_{ca \cap cb}, \\
\tg_D &=& \gamma_{DA}\vert_{db \cap dc} \cup \gamma_{DB}\vert_{dc \cap da}\cup \gamma_{DC}\vert_{da \cap db}. \label{eq:monocut4}  
\ea
$\tg_A$ is a combination of the area-minimizing cuts $\gamma_{AB}$, $\gamma_{AC}$, and $\gamma_{AD}$, such that each of the three area-minimizing cuts is restricted to the regions defined by the other two.   So $\tg_A$ separates the ER bridge into a piece containing $A$ and a piece containing all other black holes. $\tg_B$, $\tg_C$, and $\tg_D$ are defined similarly.    See Figure \ref{fig:monogamy} for an example involving four 2+1 dimensional black holes connected by a two-dimensional ER bridge.
  
The total area of the new cuts is bounded below by 
\beq\label{eq:monolower}
a(\tilde{\gamma}_A) + a(\tilde{\gamma}_B) + a(\tilde{\gamma}_C) + a(\tilde{\gamma}_D)  \geq 
\ser(A) + \ser(B) + \ser(C) + \ser(D).
\eeq
and bounded above by
\begin{align}\label{eq:monoupper}
a(\tilde{\gamma}_A) + a(\tilde{\gamma}_B) + a(\tilde{\gamma}_C) + a(\tilde{\gamma}_D) 
\leq \ser(AB) + \ser(BC) + \ser(AC),
\end{align}
Combining bounds \eqref{eq:monolower} and \eqref{eq:monoupper}, and using $\ser(D) = \ser(ABC)$, we obtain monogamy, equation \eqref{eq:mono}.

\subsection{Further inequalities}

Combining equation \eqref{eq:strong} and the triangle inequality \eqref{eq:triangle} gives an alternate version of strong subadditivity:
\beq
\ser(ABC)+\ser(B) \leq \ser(AB) +\ser(BC).
\eeq

In quantum information theory, it is sometimes convenient to work with the mutual information,
\beq 
\ier(A:B) \equiv \ser(A)+\ser(B) - \ser(AB),
\eeq
the conditional mutual information,
\beq
\ier(A:C\vert  B) \equiv \ser(AB) + \ser(BC) - \ser(B) - \ser(ABC) \geq 0,
\eeq
and the interaction information \cite{2009JHEP...03..048C,yeung1991,mcgill1954},
\ba 
\ier^3(A:B:C) &\equiv& \ser(A) +\ser(B)+\ser(C) \\ \nonumber
&-&\ser(AB) -\ser(BC) -\ser(AC) +\ser(ABC).
\ea
With these definitions, subadditivity becomes $\ier(A:B)\geq 0$, strong subadditivity becomes $\ier(A:C\vert B)>0$, and monogamy becomes $\ier^3(A:B:C)\leq 0$.

Cadney, Linden, and Winter (CLW) \cite{2005CMaPh.259..129L,2011arXiv1107.0624C} have obtained a new set of inequalities for entanglement entropy.  Strong subadditivity and monogamy together imply the CLW inequalities \cite{2013PhRvD..87d6003H}.

\section{Discussion}
\label{sec:discuss}

The entropy inequalities proved in the previous section apply to static, classical ER bridges.  The geometrical definition of entanglement entropy given by equations \eqref{eq:def1}-\eqref{eq:def2} does not apply to non-static spacetimes.  However, it could be generalized by following the example of CHEE for non-static, asymptotically AdS spacetimes \cite{2007JHEP...07..062H}.  CHEE appears to satisfy the same entropy inequalities as HEE \cite{2012JHEP...06..081C,2012arXiv1211.3494W}, so we expect that all (not necessarily static) classical ER bridges satisfy the entropy inequalities considered in section \ref{sec:inequalities}.

Quantum ER bridges have yet to be defined independently of the \ereqepr\ conjecture.  If the conjecture is correct, quantum ER bridges should satisfy subadditivity, the triangle inequality, strong subadditivity, and the CLW inequalities, as these are general properties of entanglement entropy.  However, unlike classical ER bridges, quantum ER bridges cannot be monogamous, because there are quantum states with positive interaction information.  A simple example is the four qubit pure state $\ket{GHZ_4}=(\ket{0000}+\ket{1111})/{\sqrt{2}}$ \cite{ghz89,2002PhRvA..65e2112V}, which has $I^3(A:B:C)=+1$.  If there is a quantum ER bridge corresponding to this state, it must be very unlike any classical ER bridge.

This shows that the ER bridge depends not just on the entanglement entropy, but on the pattern of entanglement.  Suppose we have $N$ copies of $\ket{GHZ_4}$.  Let the qubits of the $i$th copy be $A_i, B_i, C_i$, and $D_i$.  We could collapse all of the $A_i$ into a black hole $A$, collapse the $B_i$ into a black hole $B$, and so on.  If the number of qubits is large, then the black holes are macroscopic and have arbitrarily large entanglement entropy.  However, the interaction information of the black holes remains positive, so there is no classical ER bridge describing their entanglement.  This does not necessarily contradict \ereqepr\ because it is still possible for the entanglement between the black holes to be described by a quantum ER bridge.  This possibility cannot be ruled out without a better understanding of quantum ER bridges.  It is striking that quantum bridges do not necessarily become classical in the limit of infinite entropy.  Classical ER bridges seem to emerge when quantum correlations dominate classical correlations.  For example, Bell pairs contain purely quantum correlations.  We expect large black holes built out of Bell pairs have classical ER bridges, because all combinations of Bell pairs are monogamous.  However, the $\ket{GHZ_4}$ state contains a mix of classical and quantum correlations.  As we increase the number of $\ket{GHZ_4}$ states, the ratio of quantum to classical correlations stays fixed, the interaction information stays positive, and a classical geometry fails to emerge.

\section{Conclusions}
\label{sec:conclude}

We have shown that the entropy $\ser$ of static, classical ER bridges satisfies the usual entropy inequalities.  This is evidence for the \ereqepr\ conjecture.  The proofs apply to ER bridges in all dimensions.  We suspect the same inequalities continue to hold for non-static ER bridges, as discussed in section \ref{sec:discuss}. Our proofs are similar to the proofs of these inequalities for HEE \cite{2007PhRvD..76j6013H,2013PhRvD..87d6003H}.  However, the context is different.  We note that while HEE applies to regions on the boundary of AdS, the \ereqepr\ conjecture can apply to astrophysical black holes, in principle.  So its properties might be interesting to astrophysicists and cosmologists who are unfamiliar with AdS.  

Only special quantum states can admit a classical ER bridge description: they must have nonpositive interaction information.  In other words, classical ER requires monogamous EPR.  Large black holes with massive amounts of entanglement between them can fail to have a classical ER bridge if their entangled states have positive interaction information.  This can happen if the black holes are built out of $\ket{GHZ_4}$ states.

\section*{Acknowledgments}

We thank Max Tegmark for discussions.  R.F.P was supported in part by a Pappalardo Fellowship in Physics at MIT. 

\clearpage
\bibliographystyle{nbADS}
\bibliography{ms}

\begin{thebibliography}{10}
\ifx\href\asklfhas\newcommand{\href}[2]{#2}\fi
\ifx\arxivref\asklfhas\newcommand{\arxivref}[2]{\href{http://arxiv.org/abs/#1}{#2}}\fi
\ifx\doiref\asklfhas\newcommand{\doiref}[2]{\href{http://dx.doi.org/#1}{#2}}\fi
\raggedright
\small
\parskip 0pt

\bibitem{2010GReGr..42.2323V}
M.~{van Raamsdonk},
\textit{``{Building up spacetime with quantum entanglement}''},
\textsf{\doiref{10.1007/s10714-010-1034-0}{General~Relativity~and~Gravitation~42,~2323~(2010)}},
\texttt{\arxivref{1005.3035}{arxiv:1005.3035}}.

\bibitem{2013arXiv1306.0533M}
J.~{Maldacena} and L.~{Susskind},
\textit{``{Cool horizons for entangled black holes}''},
\texttt{\arxivref{1306.0533}{arxiv:1306.0533}}.

\bibitem{1935PhRv...48...73E}
A.~{Einstein} and N.~{Rosen},
\textit{``{The Particle Problem in the General Theory of Relativity}''},
\textsf{\doiref{10.1103/PhysRev.48.73}{Physical~Review~48,~73~(1935)}}.

\bibitem{1935PhRv...47..777E}
A.~{Einstein}, B.~{Podolsky} and N.~{Rosen},
\textit{``{Can Quantum-Mechanical Description of Physical Reality Be Considered
  Complete?}''},
\textsf{\doiref{10.1103/PhysRev.47.777}{Physical~Review~47,~777~(1935)}}.

\bibitem{1962PhRv..128..919F}
R.~W.~{Fuller} and J.~A.~{Wheeler},
\textit{``{Causality and Multiply Connected Space-Time}''},
\textsf{\doiref{10.1103/PhysRev.128.919}{Physical~Review~128,~919~(1962)}}.

\bibitem{1993PhRvL..71.1486F}
J.~L.~{Friedman}, K.~{Schleich} and D.~M.~{Witt},
\textit{``{Topological censorship}''},
\textsf{\doiref{10.1103/PhysRevLett.71.1486}{Physical~Review~Letters~71,~1486~(1993)}},
\texttt{\arxivref{gr-qc/9305017}{gr-qc/9305017}}.

\bibitem{1999PhRvD..60j4039G}
G.~J.~{Galloway}, K.~{Schleich}, D.~M.~{Witt} and E.~{Woolgar},
\textit{``{Topological censorship and higher genus black holes}''},
\textsf{\doiref{10.1103/PhysRevD.60.104039}{\prd~60,~104039~(1999)}},
\texttt{\arxivref{gr-qc/9902061}{gr-qc/9902061}}.

\bibitem{2013arXiv1307.1132J}
K.~{Jensen} and A.~{Karch},
\textit{``{The holographic dual of an EPR pair has a wormhole}''},
\texttt{\arxivref{1307.1132}{arxiv:1307.1132}}.

\bibitem{2013arXiv1307.6850S}
J.~{Sonner},
\textit{``{The ER = EPR conjecture and the Schwinger Effect}''},
\texttt{\arxivref{1307.6850}{arxiv:1307.6850}}.

\bibitem{2013arXiv1307.4706M}
D.~{Marolf} and J.~{Polchinski},
\textit{``{Gauge/Gravity Duality and the Black Hole Interior}''},
\texttt{\arxivref{1307.4706}{arxiv:1307.4706}}.

\bibitem{2013arXiv1307.1604N}
H.~{Nikolic},
\textit{``{Can a wormhole be interpreted as an EPR pair?}''},
\texttt{\arxivref{1307.1604}{arxiv:1307.1604}}.

\bibitem{2005CMaPh.259..129L}
N.~{Linden} and A.~{Winter},
\textit{``{A New Inequality for the von Neumann Entropy}''},
\textsf{\doiref{10.1007/s00220-005-1361-2}{Communications~in~Mathematical~Physics~259,~129~(2005)}},
\texttt{\arxivref{quant-ph/0406162}{quant-ph/0406162}}.

\bibitem{2011arXiv1107.0624C}
J.~{Cadney}, N.~{Linden} and A.~{Winter},
\textit{``{Infinitely many constrained inequalities for the von Neumann
  entropy}''},
\textsf{IEEE~Trans.~Inf.~Theory~58,~3657~(2012)},
\texttt{\arxivref{1107.0624}{arxiv:1107.0624}}.

\bibitem{2006PhRvL..96r1602R}
S.~{Ryu} and T.~{Takayanagi},
\textit{``{Holographic Derivation of Entanglement Entropy from the anti de
  Sitter Space/Conformal Field Theory Correspondence}''},
\textsf{\doiref{10.1103/PhysRevLett.96.181602}{Physical~Review~Letters~96,~181602~(2006)}},
\texttt{\arxivref{hep-th/0603001}{hep-th/0603001}}.

\bibitem{2006JHEP...08..045R}
S.~{Ryu} and T.~{Takayanagi},
\textit{``{Aspects of holographic entanglement entropy}''},
\textsf{\doiref{10.1088/1126-6708/2006/08/045}{Journal~of~High~Energy~Physics~8,~45~(2006)}},
\texttt{\arxivref{hep-th/0605073}{hep-th/0605073}}.

\bibitem{2007PhRvD..76j6013H}
M.~{Headrick} and T.~{Takayanagi},
\textit{``{Holographic proof of the strong subadditivity of entanglement
  entropy}''},
\textsf{\doiref{10.1103/PhysRevD.76.106013}{\prd~76,~106013~(2007)}},
\texttt{\arxivref{0704.3719}{arxiv:0704.3719}}.

\bibitem{2013PhRvD..87d6003H}
P.~{Hayden}, M.~{Headrick} and A.~{Maloney},
\textit{``{Holographic mutual information is monogamous}''},
\textsf{\doiref{10.1103/PhysRevD.87.046003}{\prd~87,~046003~(2013)}},
\texttt{\arxivref{1107.2940}{arxiv:1107.2940}}.

\bibitem{2007JHEP...07..062H}
V.~E.~{Hubeny}, M.~{Rangamani} and T.~{Takayanagi},
\textit{``{A covariant holographic entanglement entropy proposal}''},
\textsf{\doiref{10.1088/1126-6708/2007/07/062}{Journal~of~High~Energy~Physics~7,~62~(2007)}},
\texttt{\arxivref{0705.0016}{arxiv:0705.0016}}.

\bibitem{2012JHEP...06..081C}
R.~{Callan}, J.~{He} and M.~{Headrick},
\textit{``{Strong subadditivity and the covariant holographic entanglement
  entropy formula}''},
\textsf{\doiref{10.1007/JHEP06(2012)081}{Journal~of~High~Energy~Physics~6,~81~(2012)}},
\texttt{\arxivref{1204.2309}{arxiv:1204.2309}}.

\bibitem{2012arXiv1211.3494W}
A.~C.~{Wall},
\textit{``{Maximin Surfaces, and the Strong Subadditivity of the Covariant
  Holographic Entanglement Entropy}''},
\texttt{\arxivref{1211.3494}{arxiv:1211.3494}}.

\bibitem{1970CMaPh..18..160A}
H.~{Araki} and E.~H.~{Lieb},
\textit{``{Entropy inequalities}''},
\textsf{\doiref{10.1007/BF01646092}{Communications~in~Mathematical~Physics~18,~160~(1970)}}.

\bibitem{1973JMP....14.1938L}
E.~H.~{Lieb} and M.~B.~{Ruskai},
\textit{``{Proof of the strong subadditivity of quantum-mechanical entropy}''},
\textsf{\doiref{10.1063/1.1666274}{Journal~of~Mathematical~Physics~14,~1938~(1973)}}.

\bibitem{2009JHEP...03..048C}
H.~{Casini} and M.~{Huerta},
\textit{``{Remarks on the entanglement entropy for disconnected regions}''},
\textsf{\doiref{10.1088/1126-6708/2009/03/048}{Journal~of~High~Energy~Physics~3,~48~(2009)}},
\texttt{\arxivref{0812.1773}{arxiv:0812.1773}}.

\bibitem{yeung1991}
R.~W.~{Yeung},
\textit{``{A new outlook on Shannon’s information measures}''},
\textsf{IEEE~Trans.~Inf.~Theory~37,~~(1991)}.

\bibitem{mcgill1954}
W.~J.~{McGill},
\textit{``{Multivariate information transmission}''},
\textsf{Psychometrika~19,~~(1954)}.

\bibitem{ghz89}
D.~M.~{Greenberger}, M.~A.~{Horne} and A.~{Zeilinger},
\textit{``Bell's Theorem, Quantum Theory, and Conceptions of the Universe''},
Kluwer (1989),
Dordrecht,
69-72p.

\bibitem{2002PhRvA..65e2112V}
F.~{Verstraete}, J.~{Dehaene}, B.~{de Moor} and H.~{Verschelde},
\textit{``{Four qubits can be entangled in nine different ways}''},
\textsf{\doiref{10.1103/PhysRevA.65.052112}{\pra~65,~052112~(2002)}},
\texttt{\arxivref{quant-ph/0109033}{quant-ph/0109033}}.

\end{thebibliography}

\end{document}